\documentclass[osajnl,twocolumn,showpacs,superscriptaddress,10pt]{revtex4-1} 

\usepackage{amsmath,amssymb}
\usepackage{mathrsfs}
\usepackage{epstopdf}
\usepackage{graphicx}
\usepackage{bm,color,bbm}
\usepackage{natbib}
\usepackage{hyperref}
\usepackage{caption}
\usepackage{subcaption}

\newcommand{\nn}{\nonumber}

\begin{document}

\preprint{}

\title{Demonstrating Continuous Variable Einstein-Podolsky-Rosen Steering in spite of Finite Experimental Capabilities using Fano Steering Bounds}


\author{James Schneeloch}
\affiliation{Department of Physics and Astronomy, University of Rochester, Rochester, NY 14627}
\affiliation{Center for Coherence and Quantum Optics, University of Rochester, Rochester, NY 14627}

\author{Samuel H. Knarr}
\affiliation{Department of Physics and Astronomy, University of Rochester, Rochester, NY 14627}
\affiliation{Center for Coherence and Quantum Optics, University of Rochester, Rochester, NY 14627}

\author{Gregory A. Howland}
\affiliation{Air Force Research Laboratory, Rome, NY}

\author{John C. Howell}
\affiliation{Department of Physics and Astronomy, University of Rochester, Rochester, NY 14627}
\affiliation{Center for Coherence and Quantum Optics, University of Rochester, Rochester, NY 14627}


\date{\today}

\begin{abstract}
We show how one can demonstrate continuous-variable Einstein-Podolsky-Rosen (EPR) steering without needing to characterize entire measurement probability distributions. To do this, we develop a modified Fano inequality useful for discrete measurements of continuous variables, and use it to bound the conditional uncertainties in continuous-variable entropic EPR-steering inequalities. With these bounds, we show how one can hedge against experimental limitations including a finite detector size, dead space between pixels, and any such factors that impose an incomplete sampling of the true measurement probability distribution. Furthermore, we use experimental data from the position and momentum statistics of entangled photon pairs in parametric downconversion to show that this method is sufficiently sensitive for practical use.
\end{abstract}


\maketitle


\section{Introduction} 
Since its original formulation, Einstein-Podolsky-Rosen (EPR) steering has found uses beyond an intuitive witness of entanglement; it is an integral feature in certain information-theoretic quantum key distribution (QKD) protocols \cite{BranciardQKD2012}. Continuous-variable EPR-steering is especially interesting, as infinite-dimensional systems offer the possibility of very high-capacity quantum information applications including quantum superdense coding \cite{BennetSuperdense1992} and quantum teleportation. Currently, proving continuous variable EPR-steering requires enough measurements to determine the probability distributions of continuous observables (or a discretization thereof). Understandably, experimentally violating a steering inequality is subject to the caveats that experimental limitations impose. In a previous paper \cite{Schneeloch2012}, we demonstrated EPR-steering in the joint position and momentum statistics of photon pairs generated by spontaneous parametric downconversion (SPDC), subject to the limits of our detectors (i.e., that they are of finite size and do not sample 100 percent of all photon pairs). In this article, we derive a modified Fano's inequality \cite{Cover2006} useful for continuous entropies, and use it to develop steering inequalities that allow us to leverage the high measurement correlations in SPDC to demonstrate steering in spite of not having access to the complete measurement probability distribution. Indeed, the techniques discussed in this article allow one to demonstrate continuous variable EPR-steering using much fewer measurements than needed even to determine the joint probabilities that are within the range of a finite area detector. To show that these techniques are sensitive enough for practical demonstrations, we performed experimental measurements of the joint position and momentum probability distributions of down-converted photon pairs, and found that we are able to compensate for both a finite detector range, and dead space between pixels. With more efficient detectors, efficiency could also be accommodated as well.

\section{Foundations and Motivation}
EPR steering \cite{Wiseman2007} is a task between two parties (say, Alice and Bob) sharing many identical sufficiently entangled pairs of quantum systems. For each pair, Alice randomly chooses an observable and measures it on her half. Then, based on Alice's measurement outcomes, Bob performs tomography measurements on the systems he receives to determine the conditional states corresponding to Alice's measurement outcomes. Because the pairs are entangled, Alice can``steer'' the conditional state Bob measures through her choice of observable and her stated measurement outcomes (though she has no control over \emph{which} outcomes she obtains). To prove that EPR-steering is taking place, either Bob or a third party would dictate Alice's choice of observable, and then determine if EPR steering is indeed taking place by violating a sufficient criterion of EPR steering known as an EPR-steering inequality. 

EPR-steering inequalities are mathematical consequences of requiring both local causality and the completeness of a quantum-mechanical description of systems as in the original EPR scenario \cite{EPR1935}. If the effect of measurement cannot travel faster than light, then conditioning on events happening outside one's light cone cannot possibly reduce one's measurement uncertainty by more than conditioning on anything/everything inside one's light cone. Because quantum mechanics requires measurements on single systems to be constrained by uncertainty relations, locality requires that conditioning those measurements on events outside one's light cone must be constrained by those same limits as well. Understandably, many (not all \cite{chen2013allvsnothing}) EPR-steering inequalities are uncertainty relations where conditional measurement uncertainties replace standard measurement uncertainties \cite{Cavalcanti2009}. Violating an EPR-steering inequality shows that if the joint system is completely described by quantum mechanics, then there must be some correlations between spacelike-separated \footnote{Two events in spacetime are spacelike-separated if no signal traveling at or below the speed of light can travel from one event to the other. Thus, if two events are spacelike separated, they lay outside each other's light cones.} systems not reducible to information in their shared past light cones. Steering inequalities are then not as stringent as Bell inequalities, since Bell inequalities make no assumptions about a quantum description of measurement statistics.

Consider the following scenario. There are two parties, Alice and Bob, holding systems $A$ and $B$ respectively, of a joint quantum system $AB$ described by density operator $\hat{\rho}^{AB}$, and they measure position observables $\hat{x}^{A}$ and $\hat{x}^{B}$ or momentum observables $\hat{k}^{A}$ and $\hat{k}^{B}$ on their respective systems. Let us also assume that Alice and Bob's measurements are spacelike-separated. If the Universe can be described locally (as EPR assumed), and Bob's measurements are completely described by quantum mechanics, then we can conclude two things about Bob's measurements. First, his position and momentum measurements are constrained by the uncertainty relation (among others) \cite{BiałynickiBirula1975}
\begin{equation}\label{BBuncRel}
h(x^{B})+h(k^{B})\geq \log(\pi e),
\end{equation}
where $h(x^{B})$ is the continuous Shannon entropy \cite{Shannon1949,Cover2006} \footnote{Throughout this letter, all logarithms are taken to be base 2.} of Bob's position probability density. Second, conditioning on events outside his light cone (such as Alice's measurement results) reduces his measurement uncertainty by no more than by conditioning on any event(s) $\lambda$ inside his light cone;
\begin{equation}\label{LHSconstraint}
h(x^{B}|x^{A})\geq \int d\lambda \rho(\lambda) h(x^{B}|\lambda).
\end{equation}
Since this sort of relation holds true for all of Bob's measurements, we combine \eqref{BBuncRel} and \eqref{LHSconstraint} to get Walborn \emph{et.~al}'s \cite{Walborn2011} entropic steering inequality
\begin{equation}\label{CondSteerIneq}
h(x^{B}|x^{A})+h(k^{B}|k^{A})\geq \log(\pi e),
\end{equation}
where $h(x^{B}|x^{A})$ is the continuous Shannon entropy of the measurement outcomes of $\hat{x}^{B}$ conditioned on the outcomes of $\hat{x}^{A}$ \footnote{Walborn \emph{et.~al} proves this inequality more rigorously, using the non-negativity of the relative entropy, though the basic principles remain the same.}.

To demonstrate continuous variable EPR-steering with discrete measurements, it was shown in \cite{Schneeloch2012,Schneeloch2013relationship} that because the entropy of the discrete approximation of a probability density is never less than the entropy of the probability density itself; i.e.,
\begin{equation}\label{DiscApproxEnt}
H(X^{B}|X^{A})+\log(\Delta x^{B})\geq h(x^{B}|x^{A}),
\end{equation} 
violating the steering inequality,
\begin{equation}\label{DiscSteerIneq}
H(X^{B}|X^{A}) + H(K^{B}|K^{A})\geq \log\bigg(\frac{\pi e}{\Delta x^{B}\Delta k^{B}}\bigg),
\end{equation}
also violates \eqref{CondSteerIneq}. Here, $x^{A}$ and $x^{B}$ are partitioned into discrete bins of size $\Delta x^{A}$ and $\Delta x^{B}$, creating the discrete random variables $X^{A}$ and $X^{B}$ with corresponding discrete Shannon entropies $H(X^{A})$ and $H(X^{B})$ \footnote{Assuming the partitions are of the same size in both transverse directions, the steering inequality \eqref{DiscSteerIneq} in more than one dimension is altered only by multiplying the bound on the right-hand-side by the number of dimensions.}. Demonstrating continuous-variable EPR-steering this way still requires enough measurements to determine the discrete position and momentum joint probability distributions.

Though steering inequalities in the form of conditional uncertainty relations are ubiquitous \cite{Cavalcanti2009, Walborn2011, Reid1989, Schneeloch2012, Schneeloch2013}, there is no need to characterize the complete joint measurement probability distributions to demonstrate EPR-steering. If one can show that Bob's conditional measurement uncertainties violate a steering inequality by bounding them from above, then demonstrating EPR-steering only requires enough information to prove that the upper bound is sufficiently small. This was first accomplished by Cavalcanti \emph{et.~al} in 2009 \cite{Cavalcanti2009}, when they developed a steering inequality between pairs of spin-$j$ systems that only requires measuring the expected correlation between different spin components of the two particles. However, the information needed to determine an expectation value is equal to the information needed to determine the underlying probability distribution. For two-qubit systems, it was recently shown \cite{BranciardQKD2012} that one needs only half of the total joint measurement probabilities (i.e., the probabilities leading to correlated outcomes) to demonstrate EPR-steering. In what follows, we develop methods for continuous and discrete observables that also rely only on those probabilities leading to correlated outcomes. In particular, we will develop new steering bounds for discretized continuous-variable systems, (with an infinite number of possible outcomes), that can be violated with a finite number of measurements, subject to very broad constraints. We will then show with experimental data, that these bounds are sensitive enough to be of practical use.

\section{Fano Steering Bounds}
In classical information theory, Fano's inequality states that for two discrete random variables $Q^{A}$ and $Q^{B}$ which may be correlated with one another, and which share the same set of $N$ possible outcomes, the conditional Shannon entropy, $H(Q^{B}|Q^{A})$, is bounded via the following inequality:
\begin{equation}\label{FanoIneq}
h_{2}(\eta_{q}) + (1-\eta_{q}) \log(N-1) \geq H(Q^{B}|Q^{A}),
\end{equation}
where
\begin{equation}
h_{2}(\eta_{q})\equiv -\eta_{q} \log(\eta_{q}) - (1-\eta_{q})\log(1-\eta_{q}),
\end{equation}
is the binary entropy function, and $Q^{A}$ and $Q^{B}$ are discrete random variables with a finite number $N$ of outcomes. Fano's inequality \eqref{FanoIneq} gives us a finite upper bound to $H(Q^{B}|Q^{A})$, which goes to zero for an agreement probability, $\eta_{q}\equiv P(Q^{A}\!=\!Q^{B})$, approaching unity. 

The notion of using Fano's inequality to bound the conditional measurement uncertainties to witness entanglement was first discussed by Berta \emph{et.~al} \cite{Berta2010} when looking at pairs of discrete observables. They showed that sufficiently large agreement probabilities between the pairs will prove that the joint quantum system has a negative quantum conditional entropy, and hence, that it is entangled. Indeed, it is also extremely straightforward to use Fano's inequality to demonstrate EPR-steering in discrete observables, since Fano's inequality provides a useful upper bound to the conditional entropies used in sterering inequalities \cite{Schneeloch2013} for systems with a finite dimensionality $N$. What we show here is how one can use similar techniques to demonstrate EPR-steering in continuous observables (i.e. infinite-dimensional systems) where the standard Fano's inequality fails to provide a useful upper bound.

Discretized continuous observables have a (countably) infinite number of outcomes (i.e., $N\rightarrow \infty$). The only upper bound that Fano's inequality \eqref{FanoIneq} can provide in this case is that the conditional entropy is less than infinity, which is essentially useless information. However, we can use the techniques in the derivation of Fano's inequality \cite{Cover2006} to develop Fano steering bounds, which give usefully finite upper-bounds to the discrete conditional entropies in \eqref{CondSteerIneq}. We do this using only four probabilities. First, we have the two "agreement" probabilities: $\eta_{x}$, the probability that the outcome of measuring $X^{A}$ will be the same as the measurement outcome of $X^{B}$, and $\eta_{k}$, the corresponding probability for momentum. Next, we have two "domain" probabilities: $\mu_{x}$, the probability that a coincidence count will be in the $\bar{N}$-pixel domains of each detector in a joint position measurement, and $\mu_{k}$, the corresponding probability for momentum. Each agreement probability (e.g, $\eta_{x}$) is the largest sum of the diagonal elements of the joint (e.g. position) probability distribution, over all orderings of (e.g. position) outcomes. To be certain a set of measurements obtains the true agreement probability, we would have to know the complete measurement probability distribution. However, it is not necessary to know the true agreement probabilities if one has a sufficiently high lower bound for them, as one may obtain with any set of correlated measurements. This allows us to measure only those joint probabilities where we expect to see correlations, instead of having to characterize the complete joint probability distributions.

To begin deriving our modified Fano inequality, we first define two new random variables. Let $G$ be a discrete random variable with two outcomes (zero or unity). We let $G=0$ define the event $X^{A}\neq X^{B}$ (i.e., that a measurement of $X^{A}$ and $X^{B}$ will result in different outcomes). Then, we define $G=1$ to be the complementary event ``not $G=0$'', or rather, $X^{A}=X^{B}$. In addition, let $W$ be a discrete random variable with a (countably) infinite number of outcomes ($W=0,1,2,...$) dividing the sets of possible outcomes of $X^{A}$ and $X^{B}$ into domains of $\bar{N}$ pixels each (see Fig.~1). Each value of $W$ corresponds to an event where $X^{A}$ and $X^{B}$ are within a particular pair of $\bar{N}$-pixel domains. We define $W=0$ to be the event when both $X^{A}$ and $X^{B}$ are within the particular $\bar{N}$-pixel domains corresponding to the actual experimental viewing windows (i.e., the ranges of positions the detectors can measure in each arm). If $W\neq 0$, then $X^{A}$ and $X^{B}$ are within different $\bar{N}$-pixel domains, not both within the experiment's detection ranges.

\begin{figure}[t]
 \centering
\includegraphics[width=\columnwidth]{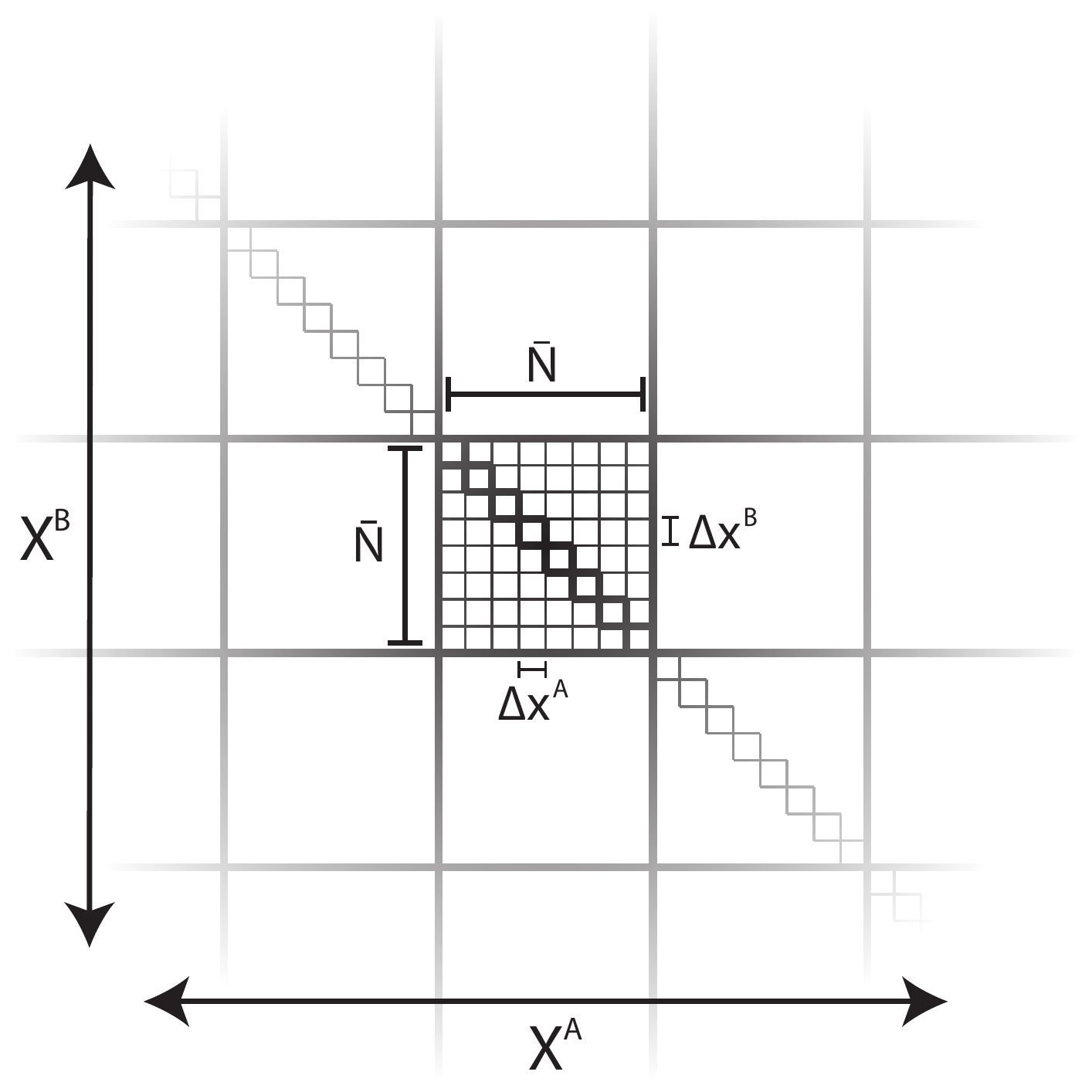}
\caption{This is a diagram of the subdivision of the joint position distribution $P(X^{A},X^{B})$ according to random variables $W$, and $G$. Each of the large squares represents a particular $\bar{N}\!\times\!\bar{N}$-pixel window for discrete position outcomes $X^{A}$ and $X^{B}$. The small squares represent joint pairs of pixels on detectors $A$ and $B$, respectively. The diagonal elements are where $G=1$. Different large squares correspond to different values of $W$.}
\end{figure}

Next, we note that adding random variables cannot reduce the entropy, so that
\begin{equation}
H(X^{B}|X^{A})\leq H(X^{B},G,W|X^{A}).
\end{equation}
In addition, removing conditioned random variables cannot reduce the entropy either, so that
\begin{equation}\label{pseudofanosum}
H(X^{B},G,W|X^{A})\leq H(G) + H(W) + H(X^{B}|G,W,X^{A}).
\end{equation}
At this point, we break $H(X^{B}|G,W,X^{A})$ up into a sum over the values of $G$:
\begin{align}\label{gsumeq}
H(X^{B}|G,W,X^{A})&= P(G=0)H(X^{B}|G=0,W,X^{A})\nn\\
&+ P(G=1)H(X^{B}|G=1,W,X^{A}).
\end{align}
When $G=1$, $X^{B}$ is wholly determined by $X^{A}$, making $H(X^{B}|G=1,W,X^{A})$ zero. Following this, we break up $H(X^{B}|G=0,W,X^{A})$ into a sum over the values of $W$:
\begin{align}\label{wsumeq}
H(X^{B}&|G=0,W,X^{A})=\nn\\
&=\sum_{i=0}^{\infty}P(W=i) H(X^{B}|G=0,W=i,X^{A}).
\end{align}
Since each window has $\bar{N}$ pixels, the entropies in the sum in \eqref{wsumeq} must be less than or equal to $\log(\bar{N}-1)$ because $G=0$. Combining the results in \eqref{pseudofanosum} through \eqref{wsumeq}, we get
\begin{equation}
H(X^{B}|X^{A})\leq  H(G) + H(W) + P(G=0) \log(\bar{N}-1).
\end{equation}
With $\eta_{x}$ already defined as the agreement probability, $P(G=0)=1-\eta_{x}$, and $H(G)=h_{2}(\eta_{x})$.

To finish the derivation, we find a finite upper limit to $H(W)$. Since $W$ is a discrete random variable with an infinite number of possible outcomes, there is no upper limit to its entropy. However, even with the extremely broad assumption that $W$ has a finite expectation value, we can construct a solid finite upper bound. We use the domain probability, $\mu_{x}=P(W=0)$ as the probability that $X^{A}$ and $X^{B}$ are within the experimental viewing window. With these two pieces of information, we know that:
\begin{equation}\label{geomMaxEnt}
H(W)\leq \frac{h_{2}(\mu_{x})}{\mu_{x}}.
\end{equation}
The reasoning works as follows. First, the maximum entropy probability distribution for a discrete random variable with finite expectation value and non-negative support is the geometric distribution. Second, the geometric distribution's entropy is a decreasing function of its maximum probability. With these two facts, $H(W)$ must be less than the entropy of a geometric distribution with maximum probability $\mu_{x}$, which is the limit in \eqref{geomMaxEnt}. In order to make this inequality especially useful, we define our \emph{measured} agreement probabilities $\bar{\eta}_{x}\equiv P(X^{A}=X^{B}|W=0)\leq \eta_{x}/\mu_{x}$ and $\bar{\eta}_{k}$ similarly for momentum, where $\bar{\eta}_{x}\rightarrow 1$ implies perfect position correlations within the experimental viewing window. By doing this, we decouple our measurement probabilities from our domain probabilities, making hedging (as discussed in the next section) considerably easier. 

With this information, we arrive at the modified Fano inequality for discretized continuous variables:
\begin{equation}\label{ModFanoIneq}
H(X^{B}|X^{A})\leq  h_{2}(\bar{\eta}_{x}\mu_{x}) + \frac{h_{2}(\mu_{x})}{\mu_{x}} + (1-\bar{\eta}_{x}\mu_{x}) \log(\bar{N}-1).
\end{equation}
Here we have made the substitution of $\bar{\eta}_{x}\mu_{x}$ for $\eta_{x}$, which makes the inequality (14) valid so long as $\bar{\eta}_{x}\mu_{x}\geq 1/2$. As we shall see, this is not a large restriction for the highly correlated systems we consider here. Note that as $\mu_{x}$ approaches unity (and 100 percent of the probability is within the viewing window), the terms accounting for a finite viewing window vanish (e.g., $\bar{\eta}_{x}\rightarrow\eta_{x}$), and we are left with the standard Fano's inequality for systems with $\bar{N}$ possible outcomes. Alternatively, using the relationship between discrete and continuous conditional entropy in \eqref{DiscApproxEnt}, we can obtain a modified Fano inequality for continuous variables.

Using this modified Fano's inequality \eqref{ModFanoIneq}, we can bound the conditional entropies in \eqref{DiscSteerIneq} to develop the steering inequality,
\begin{align} \label{FanoSteerIneq}
h_{2}&(\bar{\eta}_{x}\mu_{x})+ h_{2}(\bar{\eta}_{k}\mu_{k}) + \frac{h_{2}(\mu_{x})}{\mu_{x}} + \frac{h_{2}(\mu_{k})}{\mu_{k}}+\nn\\
&+ (2-\bar{\eta}_{x}\mu_{x}-\bar{\eta}_{k}\mu_{k})\log(\bar{N}-1)\geq \log\bigg(\frac{\pi e}{\Delta x^{B} \Delta k^{B}}\bigg),
\end{align}
where $\bar{N}$ is the total number of outcomes within the viewing window of $X^{A}$, as well as of $X^{B}$. We call these inequalities \eqref{FanoSteerIneq}, Fano steering bounds. Unlike EPR-steering inequalities involving sums or differences of observables, Fano steering bounds are solid demonstrations of EPR-steering since they require that measurements be made individually on each party. In addition, Fano steering bounds trade the ability to witness steering in many systems for the ability to witness it with much fewer measurements.

\section{Hedging with Fano Steering Bounds}
When $\mu_{x}$ and $\mu_{k}$ are unity, a sufficiently large $\bar{\eta}_{x}$ and $\bar{\eta}_{k}$ will drive the left hand side of \eqref{FanoSteerIneq} toward zero, violating the inequality. Because of this, we can use sufficiently large values of $\bar{\eta}_{x}$ and $\bar{\eta}_{k}$ to place lower limits on $\mu_{x}$ and $\mu_{k}$ above which the inequality will still be violated. This means one can effectively hedge against experimental limitations provided the data is correlated sufficiently strongly. In this way, one will be able to demonstrate EPR steering with high probability in spite of not having information about the entire measurement probability distribution.

To show how hedging with Fano steering bounds works in practice, we used an experimental setup conceptually identical to the one in \cite{HowlandPRX2013}. In our setup, we start with a 405nm OBIS laser from Coherent ${\textregistered}$. We use this laser to optically pump a BiBO nonlinear crystal cut for type-1 spontaneous parametric down-conversion of light which produces entangled pairs of photons. The downcoverted photon pairs exiting the crystal are split into signal and idler arms with a $50/50$ beamsplitter. This results in a 50 percent loss in coincident detection events, but does not alter the joint probability distributions we measure. The light from each arm passes though identical lenses, and is projected onto identical pixellated reflecting arrays, whose pixels can be oriented to reflect toward or away from a bucket detector (here, an avalanche photodiode). We used the reflecting arrays at 16x16 resolution (so that $\bar{N}=256$). Finally, the photons either hit the bucket detector or don't depending on whether they were reflected toward or away from it. When the reflecting arrays are in the image plane of the nonlinear crystal, we record the joint position distribution of the photon pairs by recording the coincidence counts from the detectors for each setting of the arrays. When the arrays are in the Fourier plane of the imaging lenses, we record the joint momentum distribution of the photons exiting the nonlinear crystal in the same way.

With the joint position and momentum distributions (within the detectors) obtained, we find the measured agreement probabilities in position and momentum, and check to see if they are sufficiently high to demonstrate steering. In our setup, we were able to obtain position and momentum agreement probabilities of $69.4\%$ and $75.1\%$, respectively. The size of the detector pixels, focal length of the imaging lens, and the central wavelength of the downcoverted light give us a value of $8.284$ bits for the right hand side of our Fano steering bound \eqref{FanoSteerIneq}. Assuming $\mu_{x}$ and $\mu_{k}$ are both 100 percent, this is easily sufficient to demonstrate EPR-steering with the Fano steering bound (as is seen by comparing our result to the red contours in Fig.~2). More importantly, we can have values of $\mu_{x}$ and $\mu_{k}$ less than unity, and still violate the steering bound. It is this that will allow us to compensate for some experimental limitations including finite detection range, dead space between pixels, and imperfect detection efficiencies.

To find a credible estimate for the domain probabilities $\mu_{x}$ and $\mu_{k}$, we fit Gaussians to the empirical probability distributions of light hitting each detector array. By comparing the integrals of these fitted Gaussians over the range of the detector, to the integral over all space, we found a minimum domain probability of 99.7$\%$ for position and 95.2$\%$ for momentum.

Real detectors have efficiencies less than 100 percent (ours are 62 percent), and there will be dead-space between pixels on the arrays. For our reflector arrays, we used two different sets for position and momentum measurements. The position reflector arrays had a fill factor \footnote{The fill factor of our reflector arrays is equal to the ratio of the area of the mirrors over the total area of the array. A fill factor of 90$\%$ means that 10 percent of the area of the array is not reflecting, being between pixels.} of 92$\%$, and the momentum reflector arrays had a fill factor of 100$\%$. While $\mu_{x}$ is the probability that the positions of the signal and idler photons will be measured to be within their respective ($\bar{N}$=pixel) detection ranges, we may also take $\mu_{x}$ simply to be the probability of a joint detection in the experiment, when set up to record the position probabilities. To accommodate efficiency and fill factors, we modify our estimated domain probabilities $\mu_{x}$ and $\mu_{k}$, multiplying them by the respective position and momentum fill factors $d_{fx}$, and $d_{fk}$, and the coincidence detection efficiency $\epsilon$. In this way, we can show that our data is sufficiently strongly correlated to demonstrate steering in spite of dead space between pixels, and a finite detector area, though not enough to account for our detection efficiencies. However, with more efficient detectors, this would be able to be accommodated as well.

To get a better idea of how well we can demonstrate steering with our Fano steering bound \eqref{FanoSteerIneq}, we show our data in Fig.~2. Here, we have plotted the difference between the left and right hand sides of our Fano steering bound as a function of the position and momentum measured agreement probabilities. This steering bound function was calculated for three situations. The red contours give the threshold for violating the Fano steering bound assuming perfect (100$\%$) domain probabilities. The central red contour gives the exact threshold, while the red contours on either side give values of five standard deviations in the bound above and below the threshold. The green contours give the threshold for violating the steering bound assuming 99.7$\%$ position and 95.2$\%$ momentum domain probabilities due to finite detector sizes. Finally, the blue contours give the violation threshold for the steering bound assuming both the (aforementioned) domain probabilities, and a 92$\%$ position fill factor, and 100$\%$ momentum fill factor. The black dot represents our measured agreement probabilities, which being beyond all three sets of contours, shows that we were able to demonstrate EPR-steering with high probability in spite of experimental limitations of a finite detection range, and dead space between pixels.

Though this method of demonstrating EPR-steering requires statistical estimation of $\mu_{x}$ and $\mu_{k}$, we note that this is a substantial improvement over prior demonstrations which were subject to the implicit assumptions that $\mu_{x}$, $\mu_{k}$, $d_{fx}$, $d_{fk}$, and $\epsilon$ are all unity.

\begin{figure}[h]
\includegraphics[width=\columnwidth]{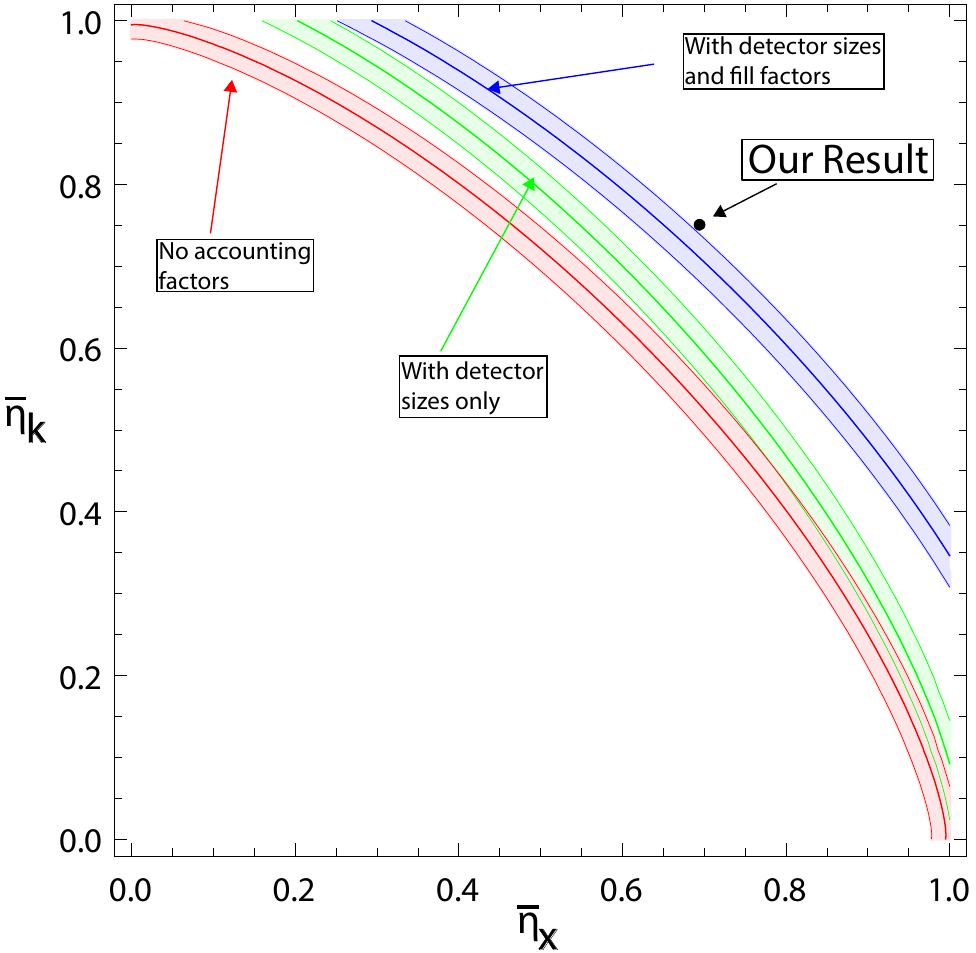}
\caption{This is a set of contour plots of the violation of our Fano steering bound as a function of the measured agreement probabilities $\bar{\eta}_{x}$ and $\bar{\eta}_{k}$. The central red contour gives the threshold for violating the Fano steering bound assuming 100$\%$ position and momentum domain probabilities, while the red contours on either side give values five standard deviations in the bound above and below this threshold. The green contours give the threshhold when we use our estimated domain probabilities given finite detector area. The blue contours give the threshhold when we use our domain probabilities accounting for both a finite detector area, and dead space between pixels. The black dot is where our measured agreement probabilities show up on the plot, indicating that we successfully demonstrate EPR-steering.}
\end{figure}

\section{Steering with fewer Measurements}
In order to test a Fano steering bound, one needs at most $2\bar{N}$ measurements to characterize the $2\bar{N}$ joint probabilities in the measured agreement probabilities $\bar{\eta}_{x}$ and $\bar{\eta}_{k}$. While this is a significant improvement to the $2N^{2}$ measurements needed to characterize both complete joint probability distributions, using the Fano steering bound can be done with perhaps much less than $2\bar{N}$ measurements since we only need to ensure sufficiently large agreement probabilities to demonstrate EPR steering. 

When it is time consuming, or otherwise expensive to characterize the entire joint probability distributions needed in traditional steering inequalities, Fano steering bounds offer a dramatic reduction in the number of required measurements to demonstrate steering by taking data only in those places where we expect to see correlations. The photon pairs in SPDC are very highly correlated in both position and momentum, so much so that most of the probability in the joint probability distribution lays in its diagonal elements (i.e. the elements contributing to the agreement probability). To obtain a large agreement probability with a relatively small number of measurements, one needs only to measure in adjacent neighborhoods where one sees coincidences, since for highly correlated photon pairs (as are known to be generated in SPDC) that is the only place where coincidences exist. Because of this, and because the elements of the agreement probability are all nonnegative, one can simply stop measuring once enough elements have been obtained to ensure that the measured agreement probabilities are sufficiently large to demonstrate steering (all other factors staying constant).

This is not, however, the only dramatic improvement in recent work to the number of measurements needed to demonstrate steering. The joint probability distributions (which give the agreement probabilities) we obtained in our setup were with the compressive sensing algorithm discussed in \cite{HowlandPRX2013}, where a uniform noise floor is removed by thresholding at 10 percent \footnote{By thresholding at 10 percent, we mean that we set all elements of the joint distribution below 10 percent of the maximum equal to zero. Then, we renormalize to get a valid probability distribution again.}. This method of acquiring the joint position and momentum probabilities as discussed in \cite{HowlandPRX2013}, offers a significant advantage over the $2N^{2}$ measurements needed to obtain the complete probability distributions via raster scanning. Indeed, this advantage is comparable to the advantage obtained by needing only to measure the band of correlated probabilities to get the agreement probabilities. The advantages in using Fano steering bounds come from not needing to characterize the entire joint probability distribution, and in being able to hedge against experimental limitations.

\section{Fano steering bounds in Continuous-Variable QKD}
Because Fano steering bounds offer the ability to demonstrate continuous-variable steering in spite of certain experimental limitations, they have a use in improving the rigor in continuous-variable quantum key distribution (QKD) schemes at the expense of a reduction in the verifiable secret key rate. As seen in \cite{nunn2013large}, the ability to violate an entropic steering inequality can be used to establish a lower bound on a one-way secret key rate between two parties.

To show how this works (also seen in \cite{nunn2013large}), we begin by using a complementary information tradeoff \cite{Renes2009}, which tells us that large correlations between Alice's and Bob's measurements limit the possible correlations between a complementary pair of measurements performed by Bob and Eve:
\begin{equation}\label{CIT}
H(X^{B}|X^{A}) + H(K^{B}|K^{E})\geq \log\bigg(\frac{\pi e}{\Delta x^{B}\Delta k^{B}}\bigg).
\end{equation}
Second, one can use the standard bound for a secret key rate $R$ with a classical adversary \cite{csiszar1978broadcast,devetak2005distillation}:
\begin{align}\label{secretkeyrate}
R&\geq H(X^{A}\!:\!X^{B})-H(X^{B}\!:\!X^{E})\nn\\
&=H(X^{B}|X^{E})-H(X^{B}|X^{A}).
\end{align}
By combining \eqref{CIT} with \eqref{secretkeyrate}, one can obtain the expression for the secret rate $R$:
\begin{equation}\label{secretKeyRete}
R\geq \log\bigg(\frac{\pi e}{\Delta x^{B}\Delta k^{B}}\bigg) -H(X^{B}|X^{A})-H(K^{B}|K^{A}),
\end{equation}
which states that $R$ is at least as large as the violation (in bits) of the entropic steering inequality \eqref{DiscSteerIneq}\cite{Schneeloch2013}. Finally, by using our continuous-variable Fano inequality \eqref{ModFanoIneq}, we can substitute the conditional entropies $H(X^{B}|X^{A})$ and $H(K^{B}|K^{A})$ with their corresponding expressions on the other sides of the Fano inequalities.

Ordinarily, one might consider the utility of Fano steering bounds to be a way of solidly demonstrating EPR steering in spite of not having access to the complete (infinite) joint probability distributions in position and momentum. Beyond steering, simply being able to bound the conditional entropies in this fashion \eqref{ModFanoIneq} allows one to make this secret key rate \eqref{secretKeyRete} much more rigorous. Without these sorts of continuous variable Fano inequalities, experimental determinations of $H(X^{B}|X^{A})$ and $H(K^{B}|K^{A})$ without information outside the range of detection is necessarily an approximation.

\section{Conclusion}
There has been much research into EPR-steering; in how to demonstrate it, and in its uses. In this paper, we showed how to demonstrate continuous variable EPR-steering with a Fano steering bound, a bound that requires much fewer measurements than needed to actually calculate the conditional measurement uncertainties in a traditional steering inequality. We accomplished this by developing a modified Fano inequality suitable for discretizations of continuous variables. Moreover, we have shown experimental data from entangled photon pairs confirming that these Fano steering bounds are of sufficient sensitivity to be of practical use. In addition, these bounds have the added utility of being able to conveniently account for sampling limitations in experimental data. Because these Fano steering bounds allow one to experimentally verify EPR-steering with given hardware limitations, we expect they will have much use in continuous-variable one-sided device-independent quantum cryptography \cite{BranciardQKD2012}, as well as other quantum information applications.

We gratefully acknowledge Daniel J. Lum's contributions in streamlining our measurement process, as well as support from DARPA DSO InPho Grant No. W911NF-10-1-0404, DARPA DSO Grant No. W31P4Q-12-1-0015, and AFOSR Grant No. FA9550-13-1-0019.

\bibliography{EPRbib12}

\end{document}